\documentclass[showpacs,showkeys,twocolumn,prl,preprintnumbers,amsmath,amssymb]{revtex4}
\newcommand{\beq}{\begin{equation}}
\newcommand{\eeq}{\end{equation}}

\newcommand{\id}{i\kern.06em\hbox{\raise.25ex\hbox{$/$}\kern-.60em$\partial$}}

\newcommand{\bs}{/\kern-.52em b}
\newcommand{\qs}{/\kern-.52em s}

\newcommand{\yp}{^{\prime}}

\newcommand{\dd}
{\kern.06em\hbox{\raise.25ex\hbox{$/$}\kern-.60em$\partial$}}

\newcommand{\vep}{\varepsilon}

\newcommand{\bi}{{\bf i}}
\newcommand{\bj}{{\bf j}}

\newcommand{\bk}{{\bf k}}

\newcommand{\JK}{J_{\rm K}}

\newcommand{\bS}{{\bf S}}

\newcommand{\bra}[1]{\langle #1|}
\newcommand{\ket}[1]{|#1\rangle}

\newcommand{\lj}{\langle}
\newcommand{\rj}{\rangle}
\DeclareMathAlphabet{\mathpzc}{OT1}{pzc}{m}{it}
\usepackage{mathrsfs}
\input amssym.def
\input amssym
\input epsf
\usepackage{graphicx}
\usepackage{dcolumn}
\usepackage{bm}
\usepackage{epsfig}
\begin{document}
\title{Ground-state properties and molecular theory of
           Curie temperature in the coherent potential approximation
           of diluted magnetic semiconductors }
\author{Sze-Shiang Feng$^\dag$, Mogus Mochena}
\affiliation
{Physics Department, Florida A \& M  University, Tallahassee, FL 32307}

\date{\today}
\begin{abstract}
Using spin-$\frac{1}{2}$ description of valence holes and Kondo coupling between local spins and carriers,
 GaAs-based  III-V diluted magnetic
semiconductors (DMS)are studied in the coherent potential
approximation(CPA). Our calculated relation of ground-state energy
and impurity magnetization shows that ferromagnetism is always
favorable at low temperatures. For very weak Kondo coupling, the
density of states (DOS) of the host semiconductor is not modified
much. Impurity band can be generated at the host band bottom only
when Kondo coupling is strong enough. Using Weiss molecular
theory, we predict a linear relation of Curie temperature with
respect to Kondo coupling and doping concentration $x$ if the hole density is proportional to $x$.
\end{abstract}
\pacs{75.30.Ds, 75.50.Pp,75.50.Dd}
\keywords{Ground-state, Curie temperature, coherent potential approximation, DMS}
\maketitle The ferromagnetism of DMS of III-V-type is not well
understood. To explain ferromagnetism in DMS, various models and
approaches have been
proposed\cite{Oiwa:1997}-\cite{Rajagopal:1998}. Though the models
differ from each other in details, they all agree that the
coupling between the carriers and local spins is of fundamental
importance. An issue of debate, however, is how the exchange
between localized spins is induced by the carriers. One model for
this induced exchange is the Ruderman-Kitttel-Kasuya-Yosida (RKKY)
interaction\cite{Oiwa:1997}\cite{Dietl:1997}. Another version
which results in conclusion equivalent to RKKY
 is the Zener model\cite{Dietl:2000&2001} which uses the fact that the valence holes are on $p$-orbitals.
 A third model is the double-exchange (DE) mechanism\cite{Akai:1998}. But this model
is inconsistent with the charge-transfer
 properties\cite{Dietl:2000&2001}.
Though RKKY can give a Curie temperature in agreement with
experiment, some argue that the RKKY model breaks down
here\cite{Konig:2000}\cite{Matsuda:2004} because the local coupling between the
carrier and the impurity spin is much larger than the Fermi energy
and can not be treated perturbatively. In dealing with the effect
of the localized spins, a key issue is whether or not randomness
should be taken into consideration. The above models are all mean
field approximations (MFA) which assume homogeneity and neglect
randomness. But DMSs are disordered systems with positional
disorder of Mn impurities. As concluded in\cite{Dietl:1997}\cite{Timm:2003}, disorder has
a substantial influence upon carrier magnetic susceptibility.
Hence, any first principle consideration should take into account
the randomness of the
impurities.\\
\indent A classic method of dealing with randomness is the
coherent potential approximation(CPA)\cite{Soven:1967} which has
been applied to
DMS\cite{Takahashi:1999&2001}-\cite{Bouzerar:2003}. Basing on the
formalism of\cite{Rangette:1973} and assuming very large local
spin $S$ while keeping the product $IS$ constant (where $I$ is the
Kondo -like interaction), \cite{Takahashi:1999&2001} obtained the
density of states and the relation between Curie temperature and
the doping concentration. Using the averaged carrier Green's
function, Bouzerar et al\cite{Bouzerar:2003} arrived at the
conclusion that the local coupling between the carrier and the
impurity spins must be intermediate in order to acquire
ferromagnetism. In this paper, we use the formalism of CPA in
\cite{Takahashi:1996} to study the ground-state properties of
III-V DMS. In contrast to \cite{Takahashi:2003}, we keep $S=5/2$
and treat the impurity spins fully quantum mechanically. Though it
is mostly accepted that the effective spin of valence holes is
$\frac{3}{2}$\cite{Linnarsson:1997}-\cite{Fiete:2003}, we describe
here the holes as spin-$\frac{1}{2}$ fermions. It is usually
believed that such a description can still catch the essential
physics. Because of spin-orbit interaction, the $p$-orbitals are
spilt (with split-off $\simeq 0.34$eV)  into a spin-$3/2$
multiplet and a spin-$1/2$ multiplet\cite{Luttinger:1955}. Using
spherical approximation, the kinetic energy of the Luttinger-Kohn
Hamiltonian \cite{Luttinger:1955} for the spin-3/2 multiplet takes
the form $\sum_\mu (\hbar^2\bk^2/2m_\mu)c^\dag_{\bk\mu}c_{\bk\mu}$
near the valence top after diagonalization, where $m_\mu=m_h\simeq
0.5m $ for $\mu=\pm 3/2$ and $m_\mu=m_\ell\simeq 0.07m $ for
$\mu=\pm 1/2$ ($m$ is the effective mass of a free hole ). The
interaction between the spin of holes and local 5/2-spins now
takes a $\bk$-dependent form $\sum_{\bk,\bk\yp}\bS\cdot c^\dag_\bk
{\bf J}(\bk,\bk\yp)c_{\bk\yp}\exp(-i(\bk-\bk\yp)\cdot{\bf R})
$\cite{Zarand:2002}. Since the DOS for parabolic band is
$g(\vep)=(1/2\pi^2 \hbar^3)(2m)^{3/2}\sqrt{\vep}$, we have the
ratio of DOS $g_h(\vep)/g_\ell(\vep)\simeq 19$ for heavy holes and
light holes, i.e., about $95\%$ of valence holes are heavy holes.
Therefore, it is a valid approximation to consider only heavy
holes. What is more, since the hole density is very small, the
Fermi wave vector is supposed to be very small and the it is thus
a good reasonable approximation to consider those $\bk$ values in
the interaction term. And this leads to the usual assumption that
the carriers are shallow holes and the coupling of the shallow
holes to the Mn$^{2+}$ can be
described by local Kondo interaction between 5/2-spins and 1/2-spins.\\
\indent  We study here one of the
most commonly studied DMS Ga$_{1-x}$Mn$_x$As, where the doping
concentration $x$ varies from 0.015 to 0.075 in region of interest
for ferromagnetism\cite{Oiwa:1997}. In Ga$_{1-x}$Mn$_x$As, ferromagnetism
was first realized at Curie temperature of 110K\cite{Dietl:1997}.
The carriers are holes originating from randomly distributed Mn.
The system is highly compensated\cite{Matsukura:1998}-\cite{Beschoten:1999} with a hole
density $p$ only around 10\% of the Mn density $x$. There are
different kinds of randomness, e.g. substitutional randomness,
interstitial randomness, antisite randomness, and directional
 randomness of impurity spin. It is commonly agreed now that interstitial Mn atoms and
 antisite As only reduce the hole densities and
 do not affect conduction of holes significantly. Therefore we consider only two
kinds of randomness, i.e., the random substitution of the Mn atoms
and the random direction of the impurity spins. The model Hamiltonian in our description is
\beq
H=\sum_{\bi,\bj,\sigma}t_{\bi\bj}c^\dag_{\bi\sigma}c_{\bj\sigma}+\sum_\bi
u_\bi
\eeq
where $u_\bi$ depends on whether $\bi$ is a Ga or
Mn site. For Ga-site
$u_\bi=u^{\text{G}}_\bi=E_{\text{G}}\sum_\sigma
c^\dag_{\bi\sigma}c_{\bi\sigma}$ ,
and for Mn-site $u_\bi=u^{\text{M}}_\bi=E_{\text{M}}\sum_\sigma
c^\dag_{\bi\sigma}c_{\bi\sigma}+\JK\bS_\bi\cdot{\bf s}_\bi$.
$\bS_\bi$ is the local spin of Mn at site $\bi$, ${\bf
s}=(1/2)c^\dag_\sigma\mbox{\boldmath$\tau$}_{\sigma\sigma\yp}c_{\sigma\yp}$
is the spin of a hole
where  $c^\dag_\sigma(c_\sigma)$ is the creation(annihilation) operators for holes
, spin indices $\sigma,\sigma\yp=\uparrow,\downarrow$, and $\mbox{\boldmath$\tau$}
=(\tau_1, \tau_2, \tau_3)$ are the three
usual Pauli matrices. $E_{\rm M}$
and $E_{\rm G}$ are the on-site energies for Ga and Mn and are assumed constant.
The hopping energy $t_{\bi\bj}=t$ if $\bi,\bj$ are nearest neighbors and zero otherwise and $\JK>0$
is the local Kondo coupling. The details of the lattice structure are not crucial in the following discussion.
  According to the general scheme of CPA, the virtual
unperturbed Hamiltonian is
\beq
\mathscr{H}(\vep)=\sum_{\sigma,\bk}(t_\bk+\Sigma_\sigma(\vep))c^\dag_{\bk\sigma}c_{\bk
\sigma}
\eeq
where $\vep$ is the Fourier frequency variable,
$\Sigma_\sigma(\vep)$ is the CPA self-energy to be determined
self-consistently and $t_\bk$ is the Fourier transformation of
$t_{\bi\bj}$. Then the relative perturbation $V$ is given by
\beq
V=H-\mathscr{H}(\vep)=\sum_\bi v_\bi
\eeq
where $v_\bi=v^{\rm G}_\bi=\sum_\sigma(E_{\text{G}}-\Sigma_\sigma)c^\dag_{\bi\sigma}c_{\bi
\sigma} $ for Ga and $v_\bi=v^{\rm
M}_\bi=\sum_\sigma(E_{\text{M}}-\Sigma_\sigma)c^\dag_{\bi\sigma}
c_{\bi \sigma}+\JK\bS_\bi\cdot{\bf s}_\bi $ for Mn. The reference
Green's function is $
\bra{\bi\sigma}\mathscr{R}(\vep)\ket{\bj\sigma\yp}=\bra{0}
c_{\bi\sigma}(\vep-\mathscr{H})^{-1}c^\dag_{\bj\sigma\yp}\ket{0} $
where $\ket{0}$ is the vacuum state of the $c$-operators, and the
associated $t$-matrices are $ t^{\rm G}_\bi=v_\bi^{\rm G}/(1-\mathscr{R}v_\bi^{\rm G})
 , t^{\rm M}_\bi=v^{\rm M}_\bi/(1-\mathscr{R}v^{\rm M}_\bi) $.  So the CPA equation
and DOS are given by
\beq
(1-x) t^{\rm G}_\bi +x\lj
t_\bi^M\rj_{\rm spin}=0
\eeq
and
\beq
g_\sigma(\vep)=-\frac{1}{\pi}{\rm Im} F_\sigma(\vep) \eeq where
$\lj\cdots\rj_{\rm spin}$ denotes average over the configurations
of impurity spins and $
F_\sigma(\vep)=\bra{\sigma\bi}\mathscr{R}\ket{\bi\sigma}=(1/N)\sum_\bk
[1/(\vep-t_\bk-\Sigma_\sigma)] $, where $N$ is the number of
lattice sites.
As usual, the spin-resolved bare DOS (for undoped GaAs) can be approximated by
the semicircle DOS
\beq
g_0(\vep)=\frac{2}{\pi\Delta}\sqrt{1-(\frac{\vep}{\Delta})^2}
\eeq
where $\Delta$ denotes the half-band width. At zero temperature, the carrier density for spin $\sigma$ can be
expressed as $
n_\sigma=\int^{\vep_F}_{-\infty}g_\sigma(\vep)d\vep,$
 where $\vep_F$ is the Fermi energy, and the total carrier density
$n=n_\uparrow+n_\downarrow$. The total electronic ground-state energy per site is $
\vep_g=\int^{\vep_F}_{-\infty}\vep
[g_\uparrow(\vep)+g_\downarrow(\vep)]d\vep $. Defining $
V_\uparrow=E_M-\Sigma_\uparrow+(\JK/2)S^z,
V_\downarrow=E_M-\Sigma_\downarrow-(\JK/2)S^z
,U_\sigma=V_\sigma-\JK/2, W_\uparrow=(1/4)\JK^2 S^-
S^+, W_\downarrow=(1/4)\JK^2 S^+ S^-$ , the CPA equations
can be written as
\begin{widetext}
\beq (1-x)\frac{E_{\rm G}-\Sigma_\uparrow}{1-F_\uparrow(E_{\rm
G}-\Sigma_\uparrow)} +x\lj[V_\uparrow(1-F_\downarrow
U_\downarrow)+F_\downarrow W_\uparrow] \frac{1} {(1-F_\uparrow
V_\uparrow)(1-F_\downarrow U_\downarrow)- F_\uparrow F_\downarrow
W_\uparrow}\rj_{\rm spin}=0\label{CPA1}
\eeq
\beq (1-x)\frac{E_{\rm
G}-\Sigma_\downarrow}{1-F_\downarrow(E_{\rm G}-\Sigma_\downarrow)}
+x\lj[V_\downarrow(1-F_\uparrow U_\uparrow)+F_\uparrow
W_\downarrow] \frac{1}{(1-F_\downarrow V_\downarrow)(1-F_\uparrow
U_\uparrow)-F_\uparrow F_\downarrow W_\downarrow}\rj_{\rm spin}=0\label{CPA2}
\eeq
\end{widetext}
These relations are given in \cite{Takahashi:1996} in another context.
For any $f(S^z)$, the spin average is given by
$\lj f(S^z)\rj_{\rm spin}=\sum_{S^z=-S}^{S}e^{\lambda S^z} f(S^z)/\sum_{S^z=-S}^{S}e^{\lambda S^z}$ where
$\lambda$ is determined by the condition
$\lj S^z\rj_{\rm spin}=m$, $m$ is the given magnetization of the impurity spins.
In our single-particle CPA, the Callen-Shtrikman relation\cite{Callen:1965} that tells
 there is a one-to-one correspondence between $m$ and $\lj (S^z)^n\rj_{\rm spin}$ for $n>1$ applies .
 Corresponding to bare DOS, we
have $ F^{(0)}(\vep)=(2/\Delta^2)(\vep-\sqrt{\vep^2-\Delta^2}) $ so $
F_\sigma(\vep)=F^{(0)}(\vep-\Sigma_\sigma) $. Solving for $\Sigma_\sigma$, we
have $ \Sigma_\sigma=\vep-(\Delta^2/4)F_\sigma-1/F_\sigma $.
Therefore CPA equations can be turned into equations for functions
$F_\sigma$ for a given $\vep$. Once $F_\sigma(\vep)$ are known, DOS $g_\sigma(\vep)$ and quantities like
Fermi energy can be
calculated. \\
\begin{figure}
\epsfxsize= 9.4truecm {\epsfbox{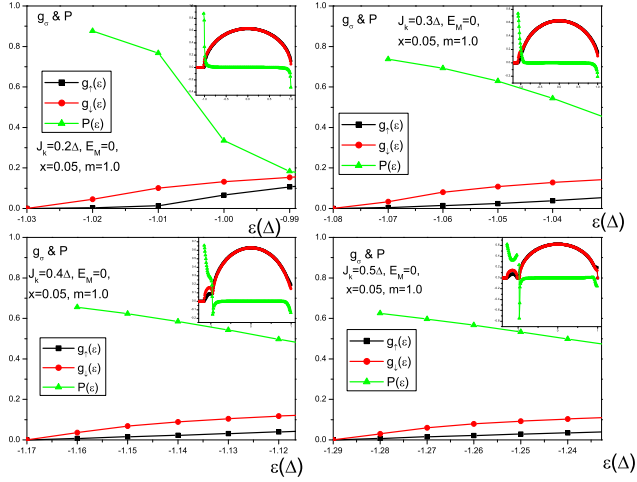}} \caption {DOS and polarization for a number of model
parameters}
\label{fig:DOS&P}
\end{figure}
\indent To solve CPA equations (\ref{CPA1}) and (\ref{CPA2}), we chose $x=0.05, p=0.1x$
 and set $E_{\rm G}=0$ since we can shift the chemical potential
without loss of physics. Energies are normalized so that
$\Delta=1$. The solution of $F_\sigma(\vep)$ is generated by iteration starting from $F^{(0)}$. First, we use the small
value of $x$ to generate $\Sigma_\sigma$ and then use the resulting $\Sigma_\sigma$ to generate
$F_\sigma$.
We calculated the DOS and the spin polarization $P(\vep)$ \cite{Zutic:2004} for model parameters
$\JK=0.1,0.2,0.3,0.4,0.5,0.6, E_{\rm M}=0,-0.2,-0.4,-0.6$
with $m=0.0.5,1.0,1.5,2.0.2.5$.
As a check for our numerical results, the sum rule $\int
^{\infty}_{-\infty} g_\sigma(z)dz=1$ is preserved and the relation
$n_\uparrow+n_\downarrow=p$ is also preserved where $n_\sigma$ is
calculated
from spin-resolved DOS respectively.
In FIG.1 we show some of our results. The curves presented in the main panels of FIG.1
are from the bottom of the band to Fermi energies at zero temperature for clarity
since this portion is important for low temperature physics. The insets outline the full behavior.
It is seen that for very weak Kondo couplings such as $\JK=0.2\Delta,0.3\Delta$, there are no impurity
bands and the DOS is not substantially
different from the bare DOS in shape. Only when Kondo coupling becomes strong enough can there be impurity bands and corresponding peaks.
In contrast to the conclusion of classical spin approximation\cite{Takahashi:2003}, the spin polarizations
are not constant up to Fermi energies.
Fig.1 shows that for $m>0$, there are always
more spin-down carriers than spin-up carriers, in compliance with the fact that the local
$p$-$d$ coupling is antiferromagnetic\cite{Dietl:2000&2001}.
 Another substantial difference between our calculated DOS
and that in\cite{Takahashi:2003} is that there is no impurity peak above the
top of the band in our calculation. So the $\vep\leftrightarrow -\vep$ symmetry of the bare DOS is broken by the
impurities.\\
\begin{figure}
\epsfxsize= 9.4truecm {\epsfbox{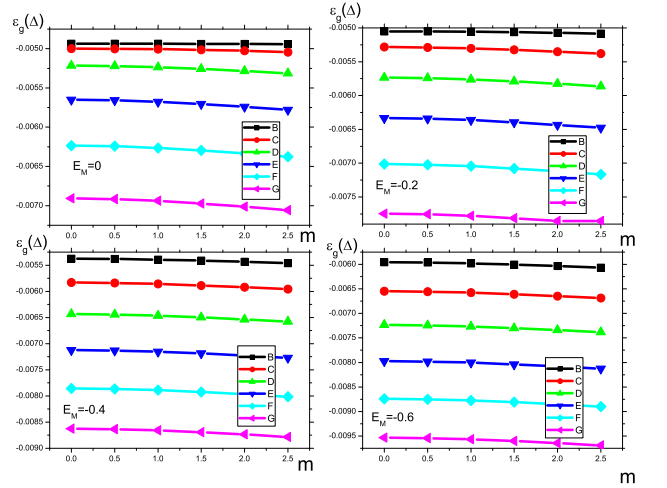}} \caption {
Relation of $\vep_g$ versus $m$.
B: $\JK=0.1$; C:$\JK=0.2$; D:$\JK=0.3$; E:$\JK=0.4$;
 F:$\JK=0.5$; G:$\JK=0.6$} \label{fig:ground}
\end{figure}
\begin{figure}
\epsfxsize= 9.4truecm {\epsfbox{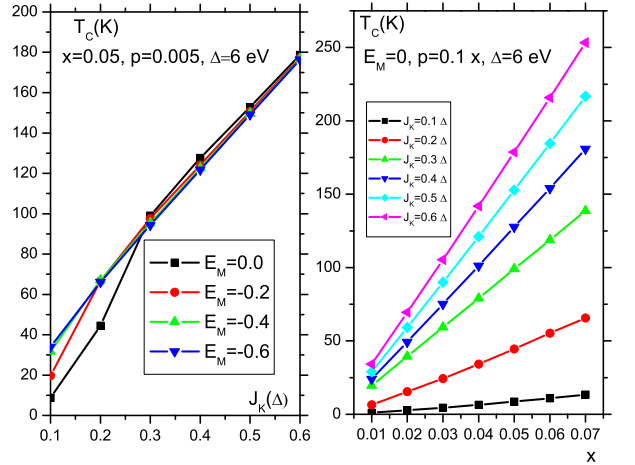}}
\caption {Theoretical estimation of Curie temperature using Weiss molecular theory} \label{fig:Tc}
\end{figure}
\indent FIG.2 shows the relation between ground-state energy
per site and the impurity magnetization $m$. In actual calculation of ground-state energy, an important issue
is the determination of the Fermi energy which is fixed by the integration of an interpolation function of the DOS. If the
interval $\delta\vep$ is not small enough, the integral of the interpolated DOS may vary significantly with the choice
of $\delta\vep$. To make it stable enough, we choose $\delta\vep=5\times 10^{-4}$ to interpolate DOS linearly.
Our results show
 that
for all the chosen values of model parameters, the ground-state energy per site always decreases, though very slowly, with increase
of impurity magnetization. Therefore CPA predicts that at very low temperatures,
 ferromagnetism is always energetically favorable for all the model parameters considered. \\
\indent In FIG.3, we show the dependence of Curie temperature on the model parameters and the doping concentration.
Weiss molecular field theory is employed as follows to calculate the Curie temperature.
Given $m$, one can calculate DOS and then $\lj s_z\rj$. So one can establish
relation $\lj s_z\rj=\lj s_z\rj(m)$.
On the other hand , given $\lj s_z\rj$, each impurity spin feels an effective field $\JK\lj s_z\rj$ and thus we have
$ m=SB_S(\beta h) $
 with $h=-\JK\lj s_z\rj(m),\beta=1/k_{\rm B}T$ and $B_S(x)$ the conventional Brillouin function.
 For very small $m$, we have
 $
\lj s_z\rj\simeq-Am
$
with $A>0$ and we have
$
\beta h\simeq\beta \JK A m
$
. So
$
B_S(\beta h)\simeq (S+1)\beta h/3
$
and thus the Curie temperature can be estimated by
$
k_{\rm B}T_{\rm C}\simeq \JK S(S+1) A/3
$
 . For small $m$,  letting
$
F_\uparrow(z)=F(z)+\psi(z)m,
F_\downarrow(z)=F(z)-\psi(z)m
$
where $F(z)$ is the paramagnetic solution,  we have
\beq
A(\JK,E_{\rm M},x,\beta)\simeq \frac{1}{\pi}\int^{\vep_F}_{-\infty} d\vep\,\, {\rm Im}\,\psi(\vep)
\eeq
where we have ignored the $\beta$-dependence. As our numerical results (not shown here) indicate,
the chemical potential
is very close to the zero temperature Fermi
energy in a wide range of temperatures
($\beta> 100$), showing that the Fermi function can be approximated by the zero temperature step
function.
Since the width of full valence band of GaAs is about $10 - 12$ eV (the width of $\Gamma_8$ band $\sim$ 4 eV)\cite{Blakemore:1982},
here we take $\Delta=6$ eV.
The left panel in FIG.3 shows the relation of $T_{\rm C}$ versus $\JK$ for different values
of $E_{\rm M}$. The curves exhibit a linear relation. For $\JK>0.3$, $T_{\rm C}$ is almost independent of $E_{\rm M}$.
The right panel shows the dependence of Curie temperature on the doping concentration for $E_{\rm M}=0$ and
various values of $\JK$. Here we still assume that the hole density $p=0.1x$. Again, the relation indicated is linear
in the range $0.01<x<0.07$. Fig.3 suggests that to reach the observed $T_{\rm C}=110$K, the value of $\JK$ needs to be
$0.3\Delta-0.4 \Delta$. For $\Delta=6$ eV, this value is much larger than the value 1 eV calculated in
\cite{Bhattacharjee:2000}. But it is still within the possible range proposed in \cite{Sanvito:2001}.
Like in the dynamical mean field study\cite{Chattopadhyay:2001} which also used semicircle
density of states and bandwidth $\sim$ 10 eV, the crucial issue in our current CPA study is the
behaviors of the Curie temperature versus the model parameters. The resulting numbers of Curire temperatures
can be scaled up or down depending on the choice of the bandwidth. \\
\indent To conclude, we summarize our results here. Using CPA and treating the impurity spins fully
quantum mechanically, we have calculated the ground-state energies of GaAs-based III-V DMS for a wide range
of model parameters. The results show that ferromagnetism is always preferable at low temperatures
. Unlike the classical treatment of the impurity spins, our approach
predicts that impurity band can arise only at bottom of the band of the host system, showing asymmetry caused by doping.
With the help of Weiss molecular theory of ferromagnetism, we obtained a linear relation of Curie temperature
with respect to Kondo coupling and doping concentration. Our results agrees with that in\cite{Schliemann:2001}.
As is known from experiments
\cite{Oiwa:1997}\cite{Matsukura:1998}, $T_{\rm C}$ increases
almost linearly with $x$ for $x<0.053$ and starts to drop when $x$
becomes larger. The contradiction might be reconciled by the
dependence on $x$ of the exchange integral (usually denoted as
$N_0\beta$). It is found in \cite{Sanvito:2001} that the absolute
value $N_0\beta$ decreases as $x$ increases, a behavior already
well known to occur in Co$_{1-x}$Mn$_x$S. Therefore, if there is a
relation like $\JK=-ax+b$ with $a$ and $b>0$, as suggested by the
data given in \cite{Sanvito:2001} ,  it can be expected that the
resulting Curie temperature may start to decrease at certain point
of doping concentration.

The authors thank Prof. Rajagopal for his helpful discussions. SF is also grateful to Prof. Takahashi
for helpful communications.
This work is supported in part by the Army High Performance Computing Research
Center(AHPCRC) under the auspices of the Department of the Army, Army Research Laboratory (ARL) under
Cooperative Agreement number DAAD19-01-2-0014.\\
$^\dag$Corresponding author, e-mail: shixiang.feng@famu.edu

\end{document}